\documentstyle[psfig]{mn}
\input{psfig.sty}

\newif\ifAMStwofonts

\title[Accretion disc coronae as magnetic reservoirs]
        {Accretion disc coronae as magnetic reservoirs
}
\author[A. Merloni \& A. C. Fabian]
        { A. Merloni and  A. C. Fabian  
\\Institute of Astronomy, Madingley Road, Cambridge, CB3 0HA
}

\date{}

\begin{document}

\maketitle

\label{firstpage}
           
\begin{abstract}
Most astrophysical sources powered by accretion onto a black hole, 
either of stellar mass or supermassive, when observed with hard X-rays show signs of
a hot Comptonizing component in the flow, the so-called {\it corona}, 
with observed temperatures and 
optical depths lying in a narrow range ($0.1 \la \tau \la 1$ and 
$ 1 \times 10^9 {\rm K} \la T \la 3 \times 10^9$ K). Here we argue that these 
facts constitute  strong supporting evidence for a magnetically-dominated 
corona. We
show that the  inferred 
thermal energy content of the corona, in all black hole systems,
is far too low to explain their observed 
hard X-ray luminosities, unless either 
the size of the corona is at least of the order of $10^3$ Schwarzschild radii, 
or the corona itself is in fact a {\it reservoir}, where the energy is mainly 
stored in the form of a magnetic field generated by a sheared 
rotator (probably the accretion disc). We briefly outline the main reasons
why the former possibility is to be discarded, and the latter preferred.
\end{abstract}

\begin{keywords}
accretion, accretion discs -- magnetic fields 
\end{keywords}

\section{Introduction}

A hard X-ray power-law is a common feature of most astrophysical systems 
 powered by an accreting black hole 
(Zdziarski 1999, and references therein).
The relative strength of the power-law component with 
respect to the quasi-blackbody one, due to thermal emission from an 
accretion disc, is used to classify the different observed spectra 
from different sources into two main states:
the {\it hard} one (when the power-law component dominates) and the {\it soft}
 one (when, on the contrary, the disk blackbody component is prominent).

The vast majority of spectral studies of galactic black hole candidates 
(GBHC) and radio-quiet Seyfert 1 
galaxies  clearly suggest that the 
primary hard X-ray continua of these sources are produced by thermal 
Comptonization (Shapiro et al. 1976; Sunyaev \& Titarchuk 1980; see Zdziarski 1999, and reference therein) in  a hot, rarefied plasma 
(hereafter, the {\it corona}) 
which probably resides where most 
of the accretion energy is released, namely in the inner part of the flow.
Furthermore, there is clear evidence that this hot, Comptonizing medium 
strongly interacts with the colder thermal component:
such an interaction is not only required to explain the ubiquitous reflection 
features in the X-ray spectra (Lightman \& White 1988; George \& Fabian 1991; Matt, Perola \& Piro 1991; Fabian et al. 2000, and reference therein), 
but could also provide the feedback 
mechanism that forces the observed values of coronal temperature and optical 
depth to lie in very narrow range for all the 
different observed sources (Haardt \& Maraschi 1991).

Theoretically, the idea that a disc configuration could explain the 
nature of the power source in black hole systems was recognized 
as early as in 1969 \cite{L69}, and the basic elements of what is still today 
considered the standard  accretion disc theory were already in place in 1973
\cite{SS73,NT73,P81}. 
Although successful in explaining many observed features of black 
hole candidates (both stellar and supermassive), standard accretion disc 
theory left unspecified the nature of both the angular momentum 
transport mechanism needed to sustain the disc in the first place, and 
the hot Comptonizing medium giving rise to the hard spectral 
component. 
 
In recent years,  it has become apparent, from both theory  
and numerical simulations
of a fully magneto-hydrodynamical accretion disc \cite{BH98}, 
that the most viable process 
for angular momentum transport involves some kind of 
turbulent magnetic viscosity. 
The dissipation of the magnetic energy built up by the magneto-rotational 
instability (MRI) in an accretion
 disc has then been shown \cite{MS2000} to produce a non uniform active corona 
which extends a few scaleheights above the disc. 

The relevance of the magnetic field for the viscosity law and the 
emission processes in  
accretion disc coronae, first introduced in a seminal paper 
by Galeev, Rosner \& Vaiana (1979),
 has already been considered in detail in Burm \& Kuperus (1988); 
Di Matteo, Celotti \& Fabian (1997); 
Di Matteo, Blackman \& Fabian (1998); Di Matteo (1998); 
Di Matteo, Celotti \& Fabian (1999), and in Wardzi\'nski and Zdziarski (2000). 
They all {\it assume} that the corona is the region 
where a significant fraction
of the total accretion power is dissipated and consequently work out the 
strength of the magnetic field, and the importance of the related 
radiative processes (cyclo-synchrotron in particular). These are in turn 
compared with the observed luminosities and spectra 
in order to assess the viability
of any such model.

The purpose of our letter, instead, is to demonstrate, with a simple 
energetics argument centered on the small thermal energy content of 
any plausible corona,  that a strong magnetic 
field in the coronal region of black hole accretion discs (both stellar 
and supermassive) is {\it required} by the existing data.   
In particular, we show that the measured thermal energy of the
Comptonizing flow is far too small to explain the observed hard X-rays 
luminosities without postulating that the flow is itself
a {\it magnetic reservoir}:  the magnetic
field dominates the energy balance in the corona and
act as an {\it in situ} reservoir of energy that powers the high energy 
emission.

\section{Energetics}

Consider a black hole of mass $M=mM_{\odot}$ and hard X-ray luminosity $L=1.5 
\times 10^{38} f_H \dot m m$ ergs s$^{-1}$, where $\dot m$ is the accretion rate 
in units of the Eddington one and $f_H$ is simply the fraction of the total
luminosity emitted in hard X-rays.

Observed galactic black hole candidates (GBHC) whose mass has been 
reliably estimated, show that, if the source is not in quiescence, 
the value of $f_H \dot m$ ranges between 0.01 and 0.1, approximately. 

To be fairly general, let us consider the thermal energy content of the 
hot electrons in a hard emitting region of size $R=rR_S$, 
where $R_S=2GM/c^2$ is the Schwarzschild radius of a black hole of mass $M$.
We have
\begin{equation}
E_{\rm th} \simeq \pi \frac{\tau}{\sigma_T} R^2 kT_e 
\simeq 5.8 \times 10^{28} \tau T_9 r^2 m^2,
\end{equation}
where $\sigma_T$ is the Thomson scattering cross section and  
$T_e=T_9 \times 10^9$ K.

Thus, for even a local equilibrium, the electrons must be heated on 
a timescale 
\begin{equation}
t_{\rm heat}=\frac{E_{\rm th}}{L} 
\simeq 3.9 \times 10^{-10} \tau T_9  r^2 m (f_H \dot m)^{-1}.
\end{equation} 
       
If the energy has to be supplied to the hot electrons from an external 
heating source, and no reservoir of stored energy is present where 
the the hot electrons are, then the light crossing time
of such region $t_{\rm cross}=R/c\simeq 10^{-5} m r$ 
(which is the shortest time-scale over which energy can be 
transferred there) has to be shorter than the heating time.
This translates into a condition on the total size of the hard X-ray emitting
region:
\begin{equation}
r > 2.5 \times 10^{4} \frac{f_H \dot m} {\tau T_9},
\label{rth}
\end{equation}
independent of the black hole mass.

For the vast majority of radio-quiet Seyfert 1s and GBHCs in their hard state, 
we deduce from observations
that the hot flow is optically thin ($0.1 \la \tau \la 1$) and has 
typical temperatures of the order of $ 1 \la T_9 \la 3$ (see e.g. 
Gierli\'nski et al. 1997; Poutanen \& Coppi 1998; Zdziarski et al. 1998; Zdziarski 1999; Petrucci et al. 2000; Done et al. 2000), implying a 
typical dimension of the Comptonizing region of the order of thousands 
Schwarzschild radii: we have, for example,  $r > 1250$ 
for the set of typical values
$\tau \simeq 0.6$, $T_9 = 1.6$ (corresponding to $kT_e \simeq 100$ keV) and 
$f_H \dot m = 0.05$. \footnote{Such a strong constraint can in principle be 
alleviated if, as is plausible, 
 the total hard X-ray emitting area is divided into a number $N$ of 
independent regions. However, the strong variability observed in these sources
can be used to set an upper limit $N \la 10$ \cite{DCF99,PF99,WZ2000}.}

Such a large coronal region is not physically plausible, 
for a number of reasons. First of all it is far too 
large to explain the fastest observed variability, both for galactic black
hole candidates (see e.g. Poutanen \& Fabian 1999; Maccarone et al. 2000) 
and for AGN (Lee et al. 2000). Strongly variable emission is clearly 
indicative of an emission cycle made of an energy storage phase followed by 
an energy release phase. 
Thus, the corona cannot be a uniform continuous medium, unless it
is geometrically thin (sheet-like; see Celotti, Fabian \& Rees 1992) 
so that the crossing time in one direction 
is orders of magnitude shorter that that in the other direction. 
Secondly, spectral evidence of a strong reflection component and  
broadened iron K$\alpha$ emission lines in Seyfert 1s 
can be explained only assuming localized hard X-rays emitting regions
shining above the inner part of the accretion disc. Also, the 
different ratios of accretion disc (blackbody-like) to hard X-ray luminosity 
observed in different sources, or in the same source at different times, 
imply that the geometry of the coronal plasma cannot be a slab one, but is 
rather made up of a number of distinct active regions \cite{HMG94}.  
Furthermore, if the external source of energy is the underlying 
accretion disc, the amount of energy deposited in the
corona is proportional to the local disc viscous power, 
which is a strong function of the radial distance and 
decays rapidly  outwards, so that almost all the energy is concentrated in the 
few inner tens of Schwarzschild radii. 
Further spectral evidence against a uniform 
corona extending out to thousands of $R_S$
for GBHC can be found in Zdziarski et al. (1998), Done \& \.Zycki (1999), Gilfanov et al. (1999).

The simplest solution to these problems is to assume that the corona is
a collection of $N$ small ($\sim$ few $R_S$, at most) active regions,
whose thermal electron energy is just a fraction of the total.
We envisage the magnetic field energy $B^2/8\pi$ as the main reservoir in the 
Comptonizing region. 
Thus, the condition $t_{\rm heat}>t_{\rm cross}$ for a magnetically-dominated 
coronal active region, translates into a condition on the magnetic 
field there (see Di Matteo, Celotti \& Fabian 1997):
\begin{equation}
B_c > \frac{5.8 \times 10^8}{r} \left(\frac{f_H \dot m}{f_{\rm cross} N m}\right)^{1/2}
{\rm G},
\end{equation}  
where we have introduced the factor 
$f_{\rm cross}=v_{\rm A}/c=B/c \sqrt(4\pi \rho)$ in 
order to take into account the fact that in a 
magnetically-dominated active region
the effective velocity for energy transfer is the Alfv\`en one, 
$v_{\rm A}$, which is,
under typical coronal conditions, unlikely to exceed the 
value $v_{\rm A}\sim 0.3 c$.

On the other hand, the strength of the magnetic field that rises buoyantly 
from the disc is in principle limited only by equipartition with
the disc pressure \cite{GRV79}. 
Following Wardzin\'ski and Zdziarski (2000) we     
calculate the maximum value of the magnetic field inside the disc (assuming 
equipartition) and obtain, for the disc region dominated by gas pressure,
\begin{equation}
B_d \simeq \frac{3.4 \times 10^8 \dot m^{2/5}}{(\alpha m)^{9/20}(1-f_H)^{1/20}} {\rm G},
\end{equation}
where $\alpha$ is the viscosity parameter, and 
\begin{equation}
B_d \simeq 5.2 \times 10^7  \left[\alpha m (1-f_H)\right]^{-1/2} {\rm G}
\end{equation}
for a disc region dominated by 
radiation pressure.

To be self-consistent, we need $B_d > B_c$,
which implies, for the two cases (gas and radiation 
pressure dominated discs), respectively
(note that, again, the limits below depend only very weakly on the black hole mass)
\begin{equation}
r > 1.7 f_H^{1/2} (1-f_H)^{1/20} \alpha^{9/20} m^{-1/20} \dot m^{1/10} 
(f_{\rm cross}N)^{-1/2},
\label{rg}
\end{equation}
and
\begin{equation}
r > 11 \left(\frac{f_H(1-f_H) \alpha \dot m}{f_{\rm cross} N}\right)^{1/2}.
\label{rr}
\end{equation}

As opposed to the case of a thermally dominated corona, eq. (\ref{rth}),
the sizes of magnetically dominated coronal region implied by eqs. (\ref{rg})
and (\ref{rr}), are perfectly consistent with the various spectral and temporal
constraint discussed above. 

\section{Discussion}

From the argument presented above, it should be quite clear that 
the inferred temperatures and optical depths of accretion disc coronae in 
black hole candidate systems demonstrate that the energy content of the 
thermal electrons is far too low to explain the hard X-ray 
luminosity of these sources.

Nevertheless,
 a number of questions still remain to be answered, which affect the 
estimates of the energy balance in the hot coronal plasma. It is 
still controversial whether the magnetic field is amplified in the disc 
up to equipartition with the gas or the total pressure, and this 
controversy also affects the nature of the disc viscosity law, with 
interesting consequences for the issue of stability 
(see e.g. Nayakshin, Rappaport \& Melia 2000).
Also uncertain is whether the magnetic energy in a reconnection site 
is mainly dissipated into electrons or protons. In the latter case,
as shown by Di Matteo, Blackman \& Fabian (1997), 
given the low Coulomb transfer
 rate between protons and electrons in the optically 
thin coronal plasma, it is 
quite likely that the corona at the equilibrium is a two-temperature flow 
\cite{JC2000,RC2000}. Thus, 
the ions could be another energy reservoir, sharing with the magnetic field 
most of the energy content in the corona \cite{DBF97} and acting as mediators between 
the magnetic field and the radiating electrons. 
However, they would also be 
likely heated 
to supervirial temperature in the inner region, evaporating the accretion 
disc and causing the the coronal flow 
to become effectively an outflow (see Meyer et al. 2000; 
Spruit \& Haardt 2000; Merloni et al., in preparation).
In general we have to consider the possibility that 
part of the energy dissipated 
via magnetic reconnection is converted into kinetic 
energy, causing bulk motion 
of the flaring material \cite{Bel99},
analogous of the Coronal Mass Ejection events
observed in the Sun \cite{Cia99,Cia2000}.  
Bulk motion may indeed be another means of storing energy within the corona.
Finally, it is not clear what is the 
fraction of non-thermal particles produced by 
the slow MHD shock associated with a magnetic reconnection site,
 and how they contribute to the observed spectral energy 
distribution \cite{PC98}.

Regardless of the above listed important open issues, 
we have shown that a strong
magnetic field is a key element for explaining
 the energetics of accretion flows
around black holes. From the spectral point of view, the presence of 
such magnetic field requires the inclusion of  cyclo-synchrotron emission 
processes in models. The relative importance of this process with respect to
inverse Comptonization and bremsstrahlung will in general depend on the geometry,
on the interaction between the two flow components and on the relativistic non-thermal
particle fraction 
\cite{CFR92,DCF97,WZ2000,MDF2000}.            

Our result emphasises the need for energy storage in the corona, 
with magnetic fields likely to be the dominant repository, although 
mildly relativistic bulk motion and a two temperature plasma may 
also share part of the energy content of the region, 
particularly once it has become active.
When the magnetic energy in a particular region has been mostly dissipated, 
the activity must switch on elsewhere, unless the geometry is 
extremely sheet-like.
Stochastic variability, as observed in both galactic black hole candidates and
AGN, is thus a likely consequence of a magnetically-dominated corona.
The picture we envisage for the accretion disc coronae is of a spread
of active regions, of which only a few are large and dominate at any given 
time. Short-lived flares may occur within an active region, which may itself
last for the duration of many such (possibly correlated) flares. 
The detailed geometry and 
motions of the X-ray emitting plasma are expected to be complicated, 
as indicated by observations of the Sun. 

\section*{Acknowledgments}
This work was done in the research network
``Accretion onto black holes, compact stars and protostars"
funded by the European Commission under contract number 
ERBFMRX-CT98-0195'. AM and ACF thank the PPARC and the Royal Society 
for support, respectively.

\bsp

\label{lastpage}

\end{document}